\documentclass[runningheads]{llncs}
\usepackage{graphicx}
\usepackage{cite}
\usepackage{amsmath,amssymb,amsfonts}
\usepackage{algorithmic}
\usepackage{graphicx}
\usepackage{xspace}
\usepackage{textcomp}
\usepackage{color}
\usepackage{morefloats}
\usepackage{textcomp}
\usepackage{listings}
\usepackage{subcaption}
\usepackage{multirow}
\usepackage{verbatim}
\usepackage{fancyvrb}
\usepackage[export]{adjustbox}
\usepackage{listings}
\definecolor{LightGray}{gray}{0.9}

\def\BibTeX{{\rm B\kern-.05em{\sc i\kern-.025em b}\kern-.08em
    T\kern-.1667em\lower.7ex\hbox{E}\kern-.125emX}}

\definecolor{blue}{rgb}{0,0,1.0}
\definecolor{darkgreen}{rgb}{0,0.44,0}
\definecolor{green}{rgb}{0,0.44,0}
\definecolor{darkred}{rgb}{0.44,0,0}
\definecolor{darkblue}{rgb}{0,0,0.64}
\definecolor{mygray}{rgb}{0.9,0.9,0.9}
\definecolor{mymauve}{rgb}{0.58,0,0.82}
\definecolor{myred}{rgb}{0.72,0.18,0.0} % red
\definecolor{mygreen}{rgb}    {0.0,0.72,0.0} % green
\definecolor{myblue}{rgb} {0.18,0.0,0.72} % blue
\definecolor{mycreme}{rgb}        {1.0,0.8,0.2} % creme

\newcommand{\gemm}{\textsc{gemm}\xspace}

\newcommand{\pe}{\mathrel{+\!\!=}}

\lstdefinelanguage{Assembler}
{morekeywords={movq,movl,imull,vsubps,vaddps,vmulps,addi,vfsub,vfadd,vfmul,vs1r, vle32,vfmv,vsetvli,ldr,fsub,fmul,fadd,str,fmov},
sensitive=false,
morecomment=[l]{//},
morecomment=[s]{/*}{*/},
morestring=[b]",
}

\begin{document}
\title{Performance Analysis of Matrix Multiplication for Deep Learning on the Edge}
%\thanks{Supported by organization x.}}
\author{%
Cristian Ramírez \inst{1}\and
Adri\'an Castell\'o \inst{1}\and
Héctor Martínez \inst{2}\and
Enrique~S.~Quintana-Ort\'{\i} \inst{1}
}
\institute{Universitat Polit\`ecnica de Val\`encia, Spain \email{crirabe@upv.es,\{adcastel,quintana\}@disca.upv.es} \and
           Universidad de Córdoba, Spain. \email{el2mapeh@uco.es}}
%
%\authorrunning{A. Castell\'o et al.}

%\author{First Author\inst{1}\orcidID{0000-1111-2222-3333} \and
%Second Author\inst{2,3}\orcidID{1111-2222-3333-4444} \and
%Third Author\inst{3}\orcidID{2222--3333-4444-5555}}
%
\authorrunning{C. Ramírez et al.}
\titlerunning{Performance Analysis of Matrix Multiplication on the Edge}
%\institute{Princeton University, Princeton NJ 08544, USA \and
%Springer Heidelberg, Tiergartenstr. 17, 69121 Heidelberg, Germany
\maketitle              % typeset the header of the contribution
%
%% Abstract 
\begin{abstract}

The devices designed for the Internet-of-Things encompass a large variety of distinct processor architectures, forming a highly heterogeneous zoo.
In order to tackle this, we employ a simulator to estimate the performance of the 
matrix-matrix multiplication (\gemm) kernel on processors designed to operate at the edge. 
Our simulator adheres to the modern implementations of \gemm, advocated by GotoBLAS2, BLIS, OpenBLAS, etc.,
to carefully account for the amount of data transfers across the memory hierarchy of different algorithmic variants of the kernel. 
%Armed with this tool, 
A small collection of experiments provide the necessary data to calibrate the simulator and deliver highly accurate estimations 
of the execution time for a given processor architecture.

\end{abstract}

%----Sections-------------------------------
\section{Introduction}

Deep learning (DL) technologies are currently being deployed at the edge
in order to improve safety and privacy, reduce the latency for the end-user, and/or decrease energy consumption~\cite{8327042,park2018deep,8675201}.
The  IoT (Internet-of-Things) appliances operating in this scenario comprise
a myriad of different processor designs, facing
limited computational and memory capacities as well as
strict restrictions in power supply and, sometimes, time-to-response.
As a consequence, the software running on these devices has to be carefully optimized. 

The general matrix-matrix multiplication (\gemm) is a key kernel for the realization
of the convolutional deep neural networks (DNNs)
employed in signal processing and computer vision, as well as for the transformers applied to natural language processing tasks~\cite{8114708}.
However, developing an efficient realization of \gemm is a time-consuming chore, 
aggravated by the heterogeneity of IoT architecture designs,
which requires a good expertise on high performance computing and computer architecture.

In this paper we contribute toward dealing with the development of optimized realizations of \gemm for IoT processors leveraging a performance simulator
to experiment with different algorithmic alternatives for this kernel, prior to actually implementing and testing them.
Our simulator, built upon the GotoBLAS2 ideas~\cite{Goto:2008:AHP} and the BLIS framework~\cite{BLIS1,BLIS4}, 
mimics the algorithm behavior in order to capture the data transfers across the 
memory hierarchy,  and requires only a few experimental data which can be collected via simple calibration experiments. 
The result delivers highly accurate estimations of the execution time on an GAP8 parallel-ultra-low power processor (PULP).

\section{Blocked Algorithms for GEMM}
\label{sec:gemm}

\subsection{The baseline algorithm for GEMM}

\begin{figure}[tbh!]
\centering
\begin{tabular}{c}
\includegraphics[width=0.7\textwidth]{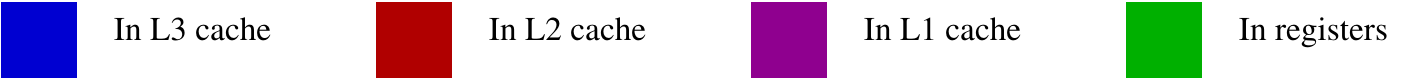}\\
\includegraphics[width=0.7\textwidth]{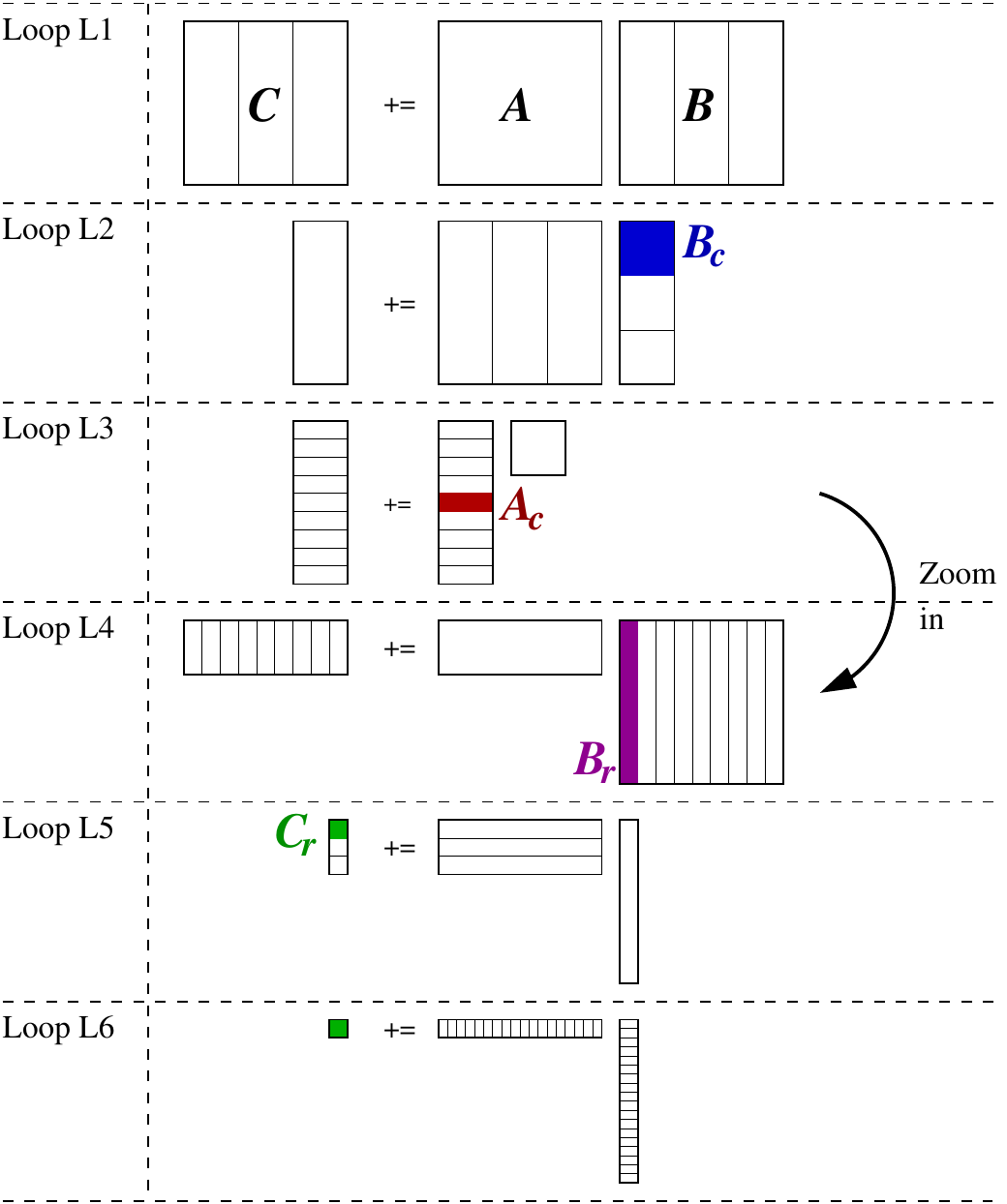}
\end{tabular}
\\
\begin{minipage}[c]{\columnwidth}
\footnotesize
\begin{verbatim}
               L1 | for ( jc=0; jc<n; jc+=nc )              
               L2 |  for ( pc=0; pc<k; pc+=kc ) {           
                  |    Bc := B(pc:pc+kc-1,jc:jc+nc-1);   (Mem->L3)
               L3 |    for ( ic=0; ic<m; ic+=mc ) {         
                  |      Ac := A(ic:ic+mc-1,pc:pc+kc-1); (Mem->L2)
               L4 |      for ( jr=0; jr<nc; jr+=nr )        
               L5 |        for ( ir=0; ir<mc; ir+=mr )      
                  |          // Micro-kernel
               L6 |          for ( pr=0; pr<kc; pr++ )      
                  |            Cc(ir:ir+mr-1,jr:jr+nr-1) (Mem->Reg)
                  |              +=  Ac(ir:ir+mr-1,pr)   (L2->Reg)
                  |              *   Bc(pr,jr:jrnr-1);   (L1->Reg)   
                  |  } }                                    
\end{verbatim}
%------------------------------------------------------------          
\end{minipage}
\caption{The baseline algorithm of \gemm. 
Here $C_c$ is a notation artifact, introduced to ease the presentation of the algorithm while 
$A_c$ and $B_c$ are actual buffers that maintain copies of certain blocks of $A$ and $B$.}
%with $C$ streamed from memory into the registers.
\label{fig:blis_family_B3A2C0}
\vspace*{4ex}
\end{figure}

Consider the \gemm
%$C := C + AB$,
%abbreviated as 
$C \pe AB$,
where the dimensions of the matrix operands
$A$,
$B$ and
$C$ are
$m \times k$,
$k \times n$ and
$m \times n$, respectively.
Many current high performance realizations of this kernel, in open-source as well as commercial linear algebra libraries,
adhere to the GotoBLAS ideas~\cite{Goto:2008:AHP} to implement it
as a collection of five nested loops around a \textit{micro-kernel} that performs a tiny \gemm. 
In rough detail, the instances of \gemm in these libraries apply tiling (blocking) to the matrix operands so that
1) a $k_c \times n_c$ block of $B$ is packed into a buffer $B_c$ that is intended to reside in the L3 cache memory; 
2) an $m_c \times k_c$ block of $A$ is packed into a buffer $A_c$ for the L2 cache memory; 
and 3) a specific $k_c \times n_r$ block of $B_c$, say $B_r$, is expected to reside in the L1 cache memory during the execution of the micro-kernel.
Furthermore, 4) the micro-kernel 
performs all the arithmetic, retrieving the data of $A_c$ from the L2 cache, $B_r$ from the L1 cache,
and $C$ directly from memory; see Figure~\ref{fig:blis_family_B3A2C0}. 
These techniques are adopted, for example, in
BLIS~\cite{BLIS1}, OpenBLAS~\cite{OpenBLAS}, AMD BLIS and, presumably, Intel MKL, among others.

\begin{figure}[tbp!]
\centering
\begin{tabular}{c}
\includegraphics[width=0.6\textwidth]{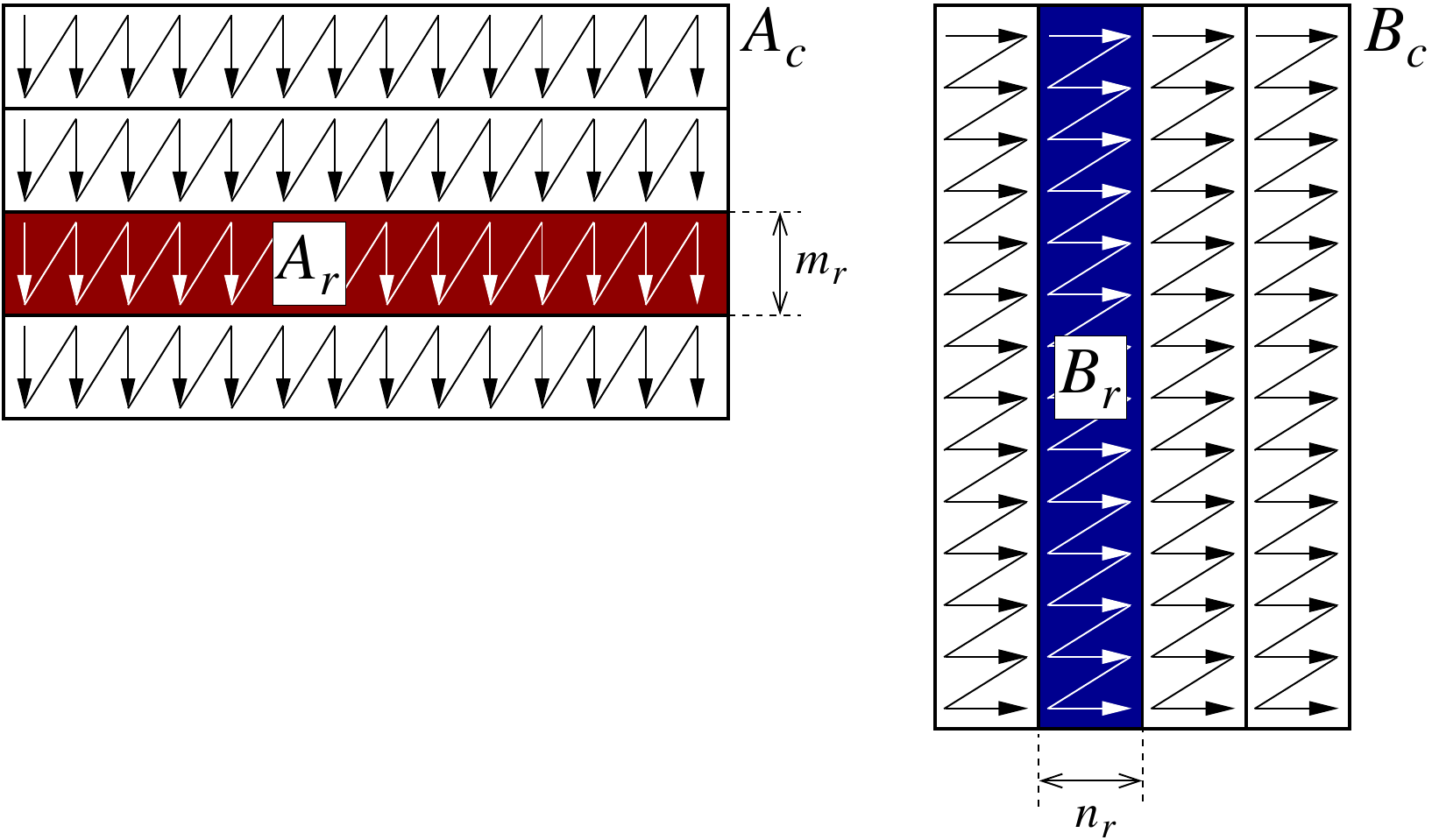}
\end{tabular}
\caption{Packing in the baseline algorithm of \gemm. Note how the entries of $A,B$ are re-organized into $A_c,B_c$ in micro-panels of
$m_r$ rows, $n_r$ columns, respectively.}
\label{fig:blis_packing}
\end{figure}

The baseline algorithm for \gemm presented in this section, hereafter referred to as B3A2C0,\footnote{%
The notation introduced in~\cite{DBLP:journals/corr/abs-1904-05717}
refers to the baseline algorithm 
as B3A2C0, where each letter denotes one of the matrix operands, and the subsequent number
%$\{0,2,3\}$, 
indicates the cache level where that operand resides (with 0 referring to the processor registers).
The same matrix operand resides in both the L1 and L3 caches.}
features a micro-kernel that comprises
a sixth loop, and is usually encoded directly in assembly (or in C with vector intrinsics).
At each iteration, this loop updates an $m_r \times n_r$ micro-tile of $C$, say $C_r$, by performing an outer product involving (part of) one row of $A_c$ and one column of $B_r$, as illustrated by loop \texttt{L6}  in Figure~\ref{fig:blis_family_B3A2C0}. The cost of loading/storing $C_r$ can be expected to 
be amortized over the $k_c$ iterations of 
this loop, as $m_r,n_r \ll k_c$ in practice.
Furthermore, a specialized packing of $A_c$ and $B_c$ ensures that their entries are retrieved with unit stride from the micro-kernel; see~Figure~\ref{fig:blis_packing}.

%\begin{figure}[tbp!]
%\centering
%\begin{tabular}{c}
%\includegraphics[width=0.5\columnwidth]{Figures/blis_microkernel_Cresident.pdf}
%\end{tabular}
%&
%\begin{tabular}{c}
%\begin{minipage}[c]{0.5\textwidth}
%\small
%\begin{tabular}{l}
             %\myfor{p_r}{k_c}{1} \\
%\hspace{3ex} \textcolor{darkgreen}{$C_c(i_r:i_r+m_r-1,j_r:j_r+n_r-1)$} \\
%\hspace{7ex}    $\pe$  \textcolor{black}{$A_c(i_r:i_r+m_r-1,p_r)$} \\
%\hspace{7ex}    ~~~~$\cdot$~\textcolor{black}{$B_c(p_r,j_r:j_r+n_r-1)$} 
%\end{tabular}
%\end{minipage}
%\end{tabular}
%\caption{Micro-kernel with $C$ resident in the processor registers.}
%\label{fig:blis_microkernel_Cresident}
%\end{figure}

\subsection{A family of algorithms for GEMM}

A different re-ordering of the \gemm loops, combined with an appropriate
selection of the loop strides, result in other variants for \gemm, which favor that the matrix blocks of $A,B,C$ reside in specific levels of the memory hierarchy, 
from the main memory to the cache(s) and processor registers. 
This was analyzed in~\cite{DBLP:journals/corr/abs-1904-05717,10.1007/11558958_30}, and
more recently, in the context of DL inference, in~\cite{CasDQ22}.

Figure~\ref{fig:blis_family_A0}
shows the algorithms for two of these variants: C3B2A0 and B3C2A0. In the former case,
1) an $m_c \times n_c$ block of $C$ is packed into a buffer $C_c$ for the L3 cache memory; 
2) a $k_c \times n_c$ block of $B$ is packed into a buffer $B_c$ for the L2 cache memory; 
and 3) an $m_r \times n_c$ block of $C_c$, say $C_r$, is intended to reside in the L1 cache memory.
In the B3C2A0 case, the roles of $C$ and $B$ are swapped.
Furthermore, 4) in both variants
the micro-kernel operates with a $m_r \times k_r$ micro-tile of $A$, streamed directly from the memory to the registers, performing
a small,  $m_r \times k_r$ matrix-vector product per iteration of Loop~\texttt{L6} ($n_c$ iterations), each involving a single column of
$C_r$ and (part of) $B_c$; see Figure~\ref{fig:blis_family_A0}.
In addition, in order to ensure accessing the entries of $C$ and $B$ with unit stride from the micro-kernel, both $C_c$ and $B_c$ are stored 
following the same pattern shown for $A_c$ in Figure~\ref{fig:blis_packing}, with $C_c$ also re-organized in micro-panels of $m_r$ rows but
$B_c$ in micro-panels of $k_r$ rows.

\begin{figure*}[tb!]
\centering
%\begin{tabular}{c}
%\includegraphics[width=0.5\textwidth]{Figures/blis_family_legend2.pdf}\\
%\includegraphics[width=0.9\textwidth]{Figures/blis_family_A0.pdf}
%\end{tabular}
%\\
\begin{minipage}[c]{\textwidth}
\footnotesize
\begin{verbatim}
            L1 | for ( jc=0; jc<n; jc+=nc )                            
            L2 |  for ( ic=0; ic<m; ic+=mc ) {                         
               |    Cc = C(ic:ic+mc-1,jc:jc+nc-1);         (Mem->L3)   
            L3 |    for ( pc=0; pc<k; pc+=kc ) {                       
               |      Bc := B(pc:pc+kc-1,jc:jc+nc-1);      (Mem->L2)   
            L4 |      for ( ir=0; ir<mc; ir+=mr )                      
            L5 |        for ( pr=0; pr<kc; pr+=kr )                    
            L6 |          for ( jr=0; jr<nc; jr++ )                    
               |            Cc(ir:ir+mr-1,jr)              (L1->Reg)   
               |              += Ac(ir:ir+mr-1,pc:pc+kr-1) (Mem->Reg)  
               |              *  Bc(pc:pc+kr-1,jr);        (L2->Reg)   
               |    }                                                  
               |    C(ic:ic+mc-1,jc:jc+nc-1) = Cc;         (L3->Mem)   
               |  }                                                    
               ------------------------------------------------------------          
            L1 | for ( jc=0; jc<n; jc+=nc )                 
            L2 |   for ( pc=0; pc<k; pc+=kc ) {              
               |     Bc := B(pc:pc+kc-1,jc:jc+nc-1);        (Mem->L3) 
            L3 |     for ( ic=0; ic<m; ic+=mc ) {            
               |       Cc := C(ic:ic+mc-1,jc:jc+nc-1);      (Mem->L2)        
            L4 |       for ( pr=0; pr<kc; pr+=kr )           
            L5 |         for ( ir=0; ir<mc; ir+=mr )         
            L6 |           for ( jr=0; jr<nc; jr++ )         
               |             Cc(ir:ir+mr-1,jr)              (L2->Reg)        
               |               += Ac(ir:ir+mr-1,pc:pc+kr-1) (Mem->Reg) 
               |               *  Bc(pc:pc+kr-1,jr);        (L1->Reg) 
               |       C(ic:ic+mc-1,jc:jc+nc-1) := Cc;      (L2->Mem)
               | } }                                        
\end{verbatim}
      %--------------------------------------------------------------------------------------------------          
\end{minipage}
\caption{Variants of the family of algorithms for \gemm with $A$ resident in the processor registers: C3B2A0 (top) and B3C2A0 (bottom).}
\label{fig:blis_family_A0}
\end{figure*}

%\begin{figure}[tbp!]
%\centering
%\begin{tabular}{c}
%\includegraphics[width=0.6\columnwidth]{Figures/blis_microkernel_Aresident.pdf}
%\end{tabular}
%&
%\begin{tabular}{c}
%\begin{minipage}[c]{0.5\textwidth}
%\small
%\begin{tabular}{l}
             %\myfor{j_r}{n_c}{1} \\
%\hspace{3ex}             \textcolor{black}{$C_c(i_r:i_r+m_r-1,j_r)$} \\
%\hspace{7ex}                    $\mathrel{+}=$    \textcolor{darkgreen}{$A_c(i_r:i_r+m_r-1,p_r:p_r+k_r-1)$} \\
%\hspace{7ex}                    ~~~~$\cdot$        ~\textcolor{black}{$B_c(p_r:p_r+k_r-1,j_r)$} \\
%\end{tabular}
%\end{minipage}
%\end{tabular}
%\caption{Micro-kernel with $A$ resident in the processor registers.}
%\label{fig:blis_microkernel_Aresident}
%\end{figure}

To close this section, we note that swapping the roles of $A$ and $B$ in the three previous algorithms, yields
three alternative variants:
A3B2C0, C3A2B0, A3C2B0~\cite{CasDQ22}. However, given the symmetric role of the input operands of \gemm ($A,B$), these other
variants present no significant differences from the point of view of the performance model proposed in this work and, therefore, we do not consider 
in the following.

\section{A Performance Simulator for GEMM Algorithms}
\label{sec:model}

\subsection{IoT architecture model}

\newcommand{\reg}{\textsc{R}\xspace}
\newcommand{\lone}{\textsc{L1}\xspace}
\newcommand{\ltwo}{\textsc{L2}\xspace}
\newcommand{\mem}{\textsc{M}\xspace}
\newcommand{\tr}[1]{T_{#1}\xspace}

We make the following considerations with respect to the target IoT processor:
\begin{itemize}
\item The processor is equipped with a single core, with a SIMD (single instruction multiple data) arithmetic units
      capable of working with 32 vector registers of width 32 bits (4 INT8 numbers).
\item The memory comprises four levels, from fastest/smallest to slowest/largest referred to as
\reg (for processor registers),
\lone, \ltwo, and \mem (for main memory).
\item There is a strict control of the data transfers between memory levels. The \lone and \ltwo 
      levels can thus be viewed as ``scratchpad'' memories
      instead of conventional caches.
\item The capacity of each memory level will be denoted as $C_{L}$,
      with $L$ denoting the corresponding level.
\item The transfer rates between two levels will be referred to as $\tr{O,D}$, with the subindices $O/D$ 
      specifying the origin/destination memory levels.
\end{itemize}
From the point of view of the algorithms, for simplicity we assume that computation is not overlapped with data transfers 
involving the scratchpad memories. 

\subsection{Validation}

\begin{table}[htb!]
\begin{center}
\renewcommand{\arraystretch}{1.2}
\setlength{\tabcolsep}{6pt}
\begin{tabular}{|l|l||l|l|l|l|}
\hline
                  & Transfer           & Mbytes/s   & B3A2C0          & C3B2A0          & B3C2A0                   
\\ \hline \hline
   Packing        & $\tr{\mem,\mem}$   & 1.62E$+$00 & $B$ to $B_c$    & $C$ to $C_c$    & $B$ to $B_c$        
\\ Packing         & $\tr{\mem,\ltwo}$  & 5.30E$-$01 & $A$ to $A_c$    & $B$ to $B_c$    & $C$ to $C_c$        
\\ Unpacking      & $\tr{\ltwo,\mem}$  & 6.54E$-$01 & --              & --              & $C_c$ to $C$      
\\ \hline
   Copy           & $\tr{\mem,\lone}$  & 8.81E$+$00 & $B_c$ to $B_r$  & $C_c$ to $C_r$  & $B_c$ to $B_r$      
\\ \hline
   Stream from    & $\tr{\mem,\reg}$   & 4.87E$-$01 & $C$ to reg.     & $A$ to reg.     & $A$ to reg.  
\\ micro-         & $\tr{\lone,\reg}$  & 1.78E+02   & $B_r$ to reg.   & $C_r$ to reg.   & $B_r$ to reg. 
\\ kernel         & $\tr{\ltwo,\reg}$  & 7.18E+00   & $A_c$ to reg.   & $B_c$ to reg.   & $C_c$ to reg. 
\\  \hline
\end{tabular}
\end{center}
\vspace*{-2ex}
\caption{Transfers rates in the GAP8 FC. The packing/unpacking rates (three first rows) were measured when transferring chunks of $r=4$ elements at a time.}
\label{tab:calibration}         
\end{table}

~\\
\noindent
\textbf{Hardware platform.}
For the validation of our performance simulator, in this work we target
the GAP8 PULP, from GreenWaves Technologies. This
system 
comprises 1) a fabric controller (FC) core for control, communications, 
and security functions;
2) a cluster of 8 cores designed for the execution of parallel algorithms;
and 3) a specialized %Convolutional Neural Network 
accelerator (HWCE).
All these components share the same 512-KB L2 \textit{memory area} (MA).
Furthermore, the FC has a 16-KB L1 MA while the cluster cores and HWCE share a 64-KB multi-banked TCDM L1 
(data/instruction) MA.
Several DMA (direct memory access) units allow fast transfers between MAs. 
%The layout of the GAP8 platform is illustrated in Figure~\ref{fig:gap8}.
The banks of the shared L1 MA can be accessed from the cluster cores in a single cycle. 
In comparison, accessing data in external MAs (referred to as L3 memory,)
incurs a very high cost and, therefore, should be avoided whenever possible. 
The GAP8 relies on DMA units to
transfer data to/from peripherals and in between the internal L1 and L2 MAs, which can be viewed as ``scratchpads''. 
The DMA unit is used to transfer data to/from peripherals, including the L3 memory. 

Following our assumptions on the IoT processor, we only target the FC core, and associated MAs, 
for the validation and experimentation in the remainder of the paper.
Repeating the analysis for the GAP8 cluster, using a multi-threaded version of \gemm, is left as part of future work.

~\\
\noindent
\textbf{Calibration.}
We conducted a series of experiments to estimate the data transfer rates between the MAs in the GAP8 FC, 
with the results offered in Table~\ref{tab:calibration}. 
The first block-row there comprises the packing/unpacking operations associated with blocking (tiling) and are performed
by the three outermost loops of the algorithms. They all
involve the L3 MA (\mem in the model), and the results were obtained using DMA programmed transfers of $r=4$ elements ``at a time''. This type of calibration is required
because packing/unpacking the matrix operands into their corresponding buffers, requires a reorganization that copies the data in ``chunks'' of $r$ consecutive
elements in memory; see Figure~\ref{fig:blis_packing}.
We could also verify that, when multiplying $r$ by a factor $s$, the transfer rate also increased in the same proportion.
For example, for algorithm B3A2C0, $B$ is packed into the buffer $B_c$ taking into account the 
dimension $n_r=4$ of the micro-kernel, 
and proceeds at a rate of  $1.62$ MBytes/s. If the micro-kernel for this algorithm is modified to use $n_r=8$, 
we experimentally observed that the rate was doubled, to $3.24$ MBytes/s.
Our simulator takes this consideration into account.

The second block-row in the table (consisting of a single row) corresponds to the copy between the L3 and L1 MAs. This copy is implicit in the case of
the conventional \gemm algorithms, which assume a cache system (and therefore, they do not appear reflected in the formulation of the algorithms), 
but they need to be explicitly programmed in the case of scratchpads.

The third block-row of results are for the data streaming performed from inside the micro-kernel.

A separate experiment with a micro-kernel designed for the GAP8 FC, with $A$ resident in the processor registers
and the two other operands placed in the proper MAs, showed an arithmetic performance
of 5.64 billions of INT8 arithmetic operations per second (INT8 GOPS).

\begin{figure}[tbh!]
\centering % \vspace{0.5cm}
\includegraphics[angle=-90,width=0.7\textwidth]{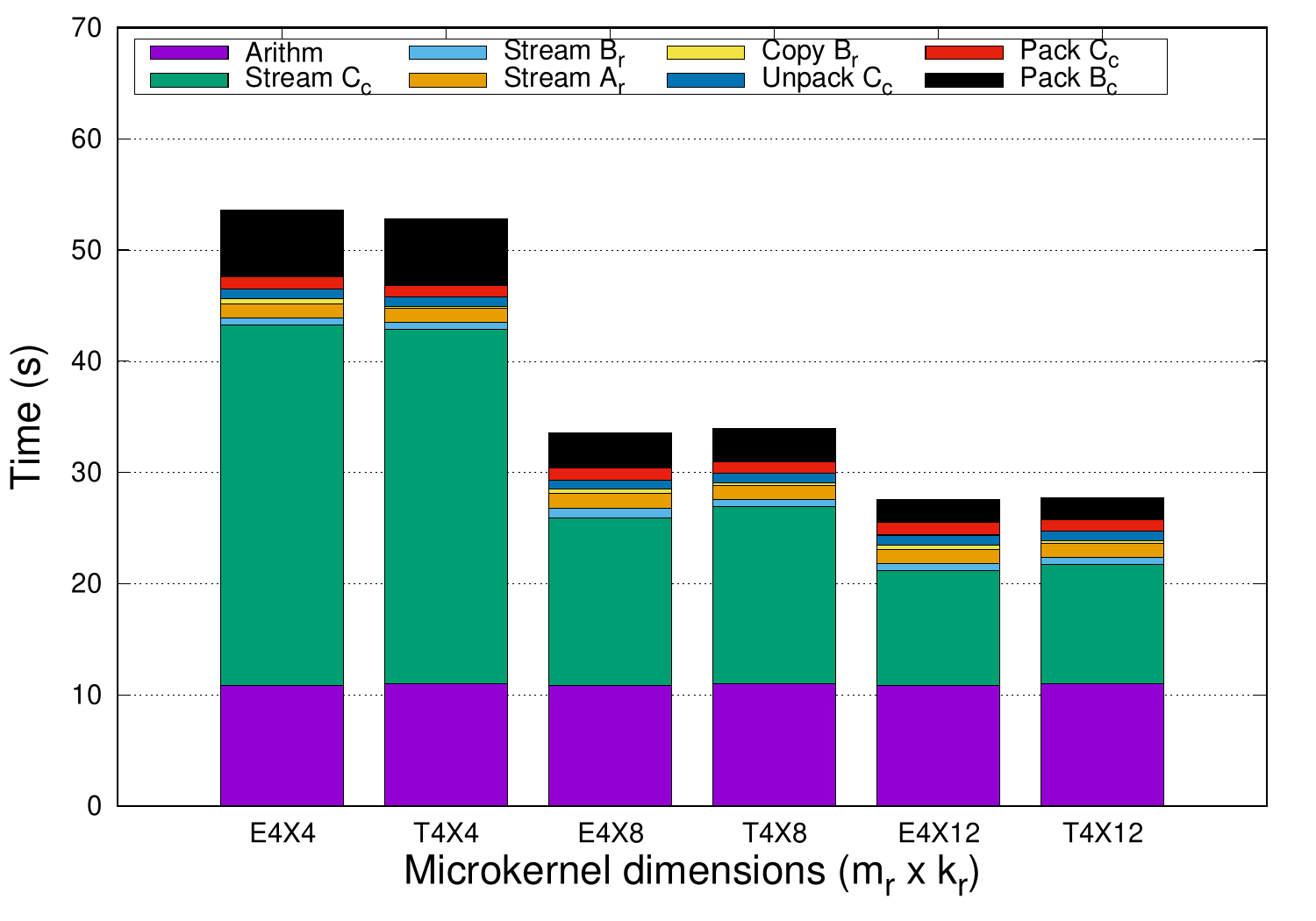}
\vspace*{-2ex}
\caption{Distribution of costs among the different components of the B3C2A0 algorithm
         using micro-kernels of dimension 
         $4\times 4$,
         $4\times 8$, and
         $4\times 12$.
%\textcolor{red}{Quitar theoretical analysis para micro-kernel $4\times 16$. Los colores de esta figura y las siguientes deberían 
%coincidir.}
The labels starting with ``E'' and ``T'' below each bar distinguish between results from experimentation and the simulator, respectively.}
\label{fig:validation}
\end{figure}

~\\
\noindent
\textbf{Validation.}
We next leveraged
our implementation of the C3B2A0 algorithm for the GAP8 FC described in~\cite{RamCQ22}
in order to assess the accuracy of our simulator.
For this purpose,
we selected a \gemm of moderate dimensions:
$m,n,k=256, 784, 2304$.
(These particular dimensions were chosen because they arise when applying the lowering approach~\cite{8114708} 
to transform the convolution operator in layer \#10 of MobileNetV1 DNN
into a \gemm.)
Once we fixed the micro-kernel dimension ($m_r \times k_r$, for this particular variant),
we then set the 
scratchpad configuration parameters ($m_c, n_c, k_c$)
so that $C_r,B_c$ respectively maximize the occupancy of the L1, L2 MAs of the GAP8 FC.

Figure~\ref{fig:validation} shows that 
the simulator, tuned with the calibrated transfer and arithmetic rates, estimates the execution time of the
actual implementation remarkably well. Overall, the relative errors of the simulator in all these tests remained below 2\%.

\section{Performance Analysis}
\label{sec:analysis}

As argued in the introduction of this paper, the ultimate goal
of our performance simulator for \gemm is to experiment with different algorithmic alternatives for the kernel, prior to 
going through the effort of implementing and testing any of them on a specific IoT processor.

\begin{figure}[tbh!]
%\ContinuedFloat
%\begin{subfigure}{\textwidth}
\centering
    %\hspace*{-3cm}
\includegraphics[angle=-90,width=1\textwidth]{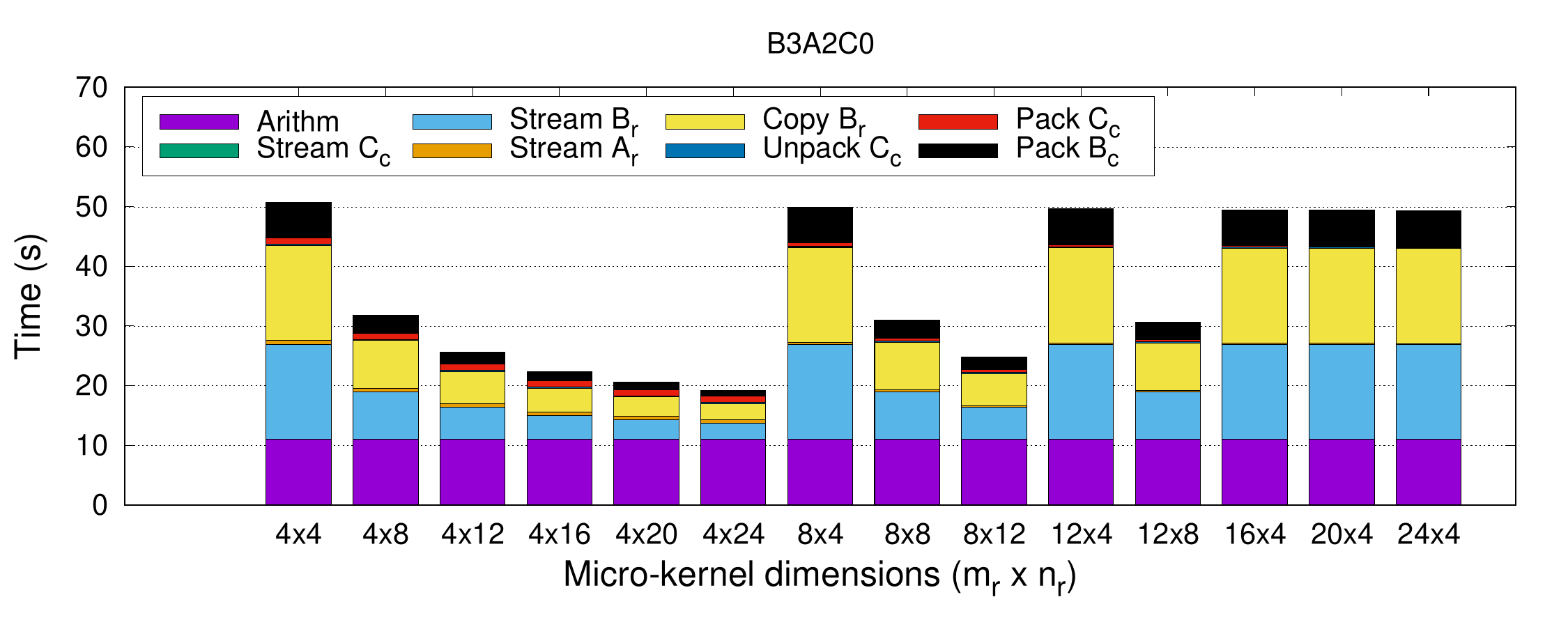}
\includegraphics[angle=-90,width=1\textwidth]{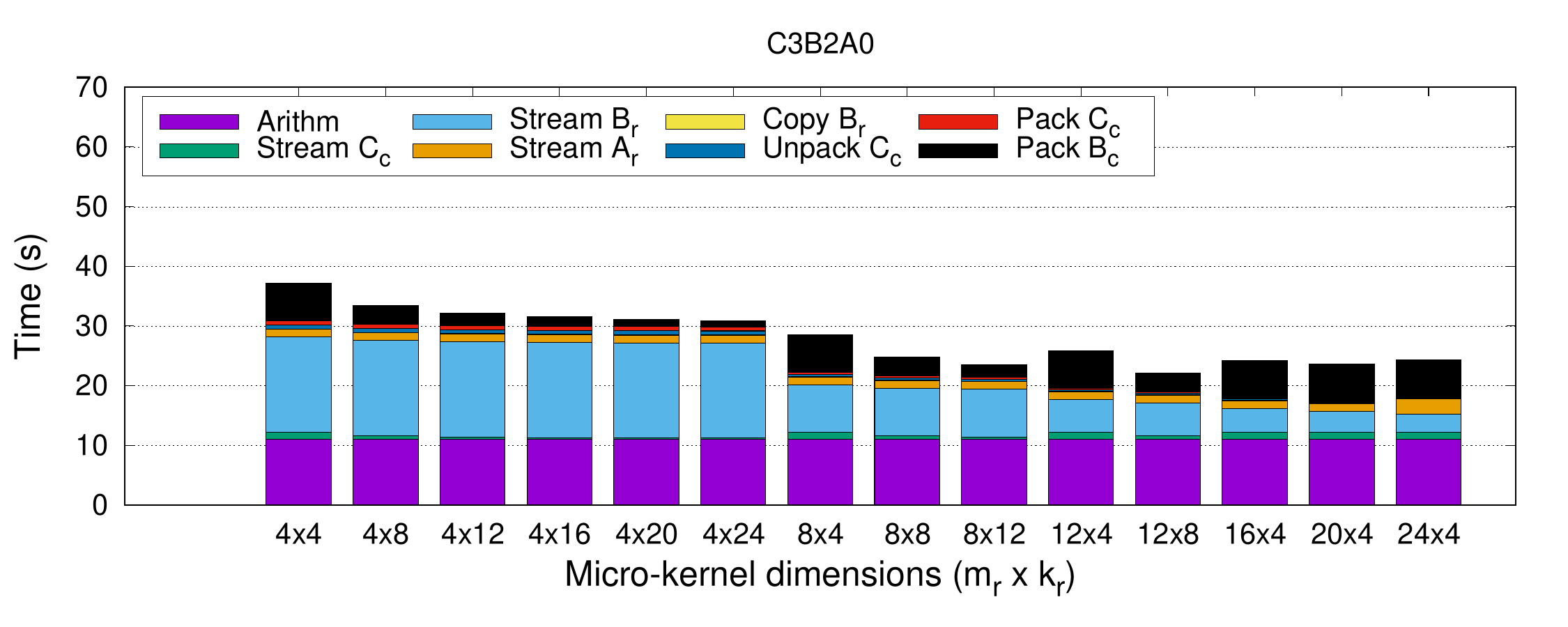}
\includegraphics[angle=-90,width=1\textwidth]{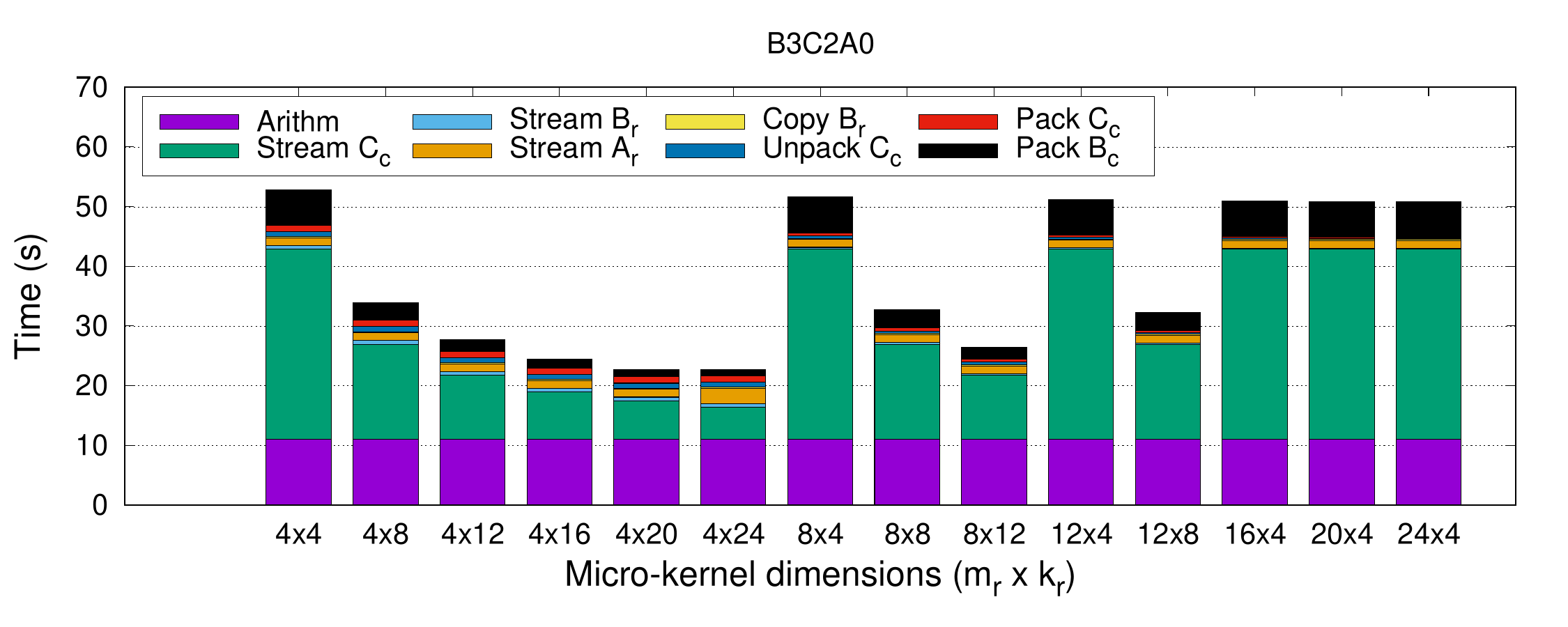}    %\caption{B3A2C0.}
    %\label{fig:second}
%\end{subfigure}
%\begin{subfigure}{\textwidth}
    %\centering
    %\hspace*{-3cm}
    %\caption{B3C2A0.}
    %\label{fig:third}
%\end{subfigure}
%\begin{subfigure}{\textwidth}
    %\centering
    %\hspace*{-3cm}
    %\caption{C3B2A0.}
    %\label{fig:third}
%\end{subfigure}
\caption{Execution time of the three algorithms for the \gemm in layer \#10 of MobileNetV1 
         estimated using the performance simulator calibrated for the GAP8.}
\label{fig:simulator}
\end{figure}

In this section we evaluate the three algorithmic variants for \gemm discussed earlier:
B3A2C0, C3B2A0 and B3C2A0, comparing their estimated performance as a function of the dimension
of the internal micro-kernel 
($m_r \times n_r$ for the first variant; and
 $m_r \times k_r$ for last two),
 and initially leveraging the same problem case from the previous section:
 $m,n,k=256, 784, 2304$.
 The size of the selected micro-kernels was determined following the assumptions on the width of the SIMD arithmetic unit (32 bits) 
 and number of vector registers (32) made in Section~\ref{sec:model}.
 
Figure~\ref{fig:simulator} shows the 
distribution of the arithmetic and data/transfer costs, for the three variants, using the performance simulator calibrated for
the GAP8 platform. An assumption of our basic simulator is that the arithmetic rate is independent of the micro-kernel dimension and
this results in all cases reporting the same cost due to arithmetic. (This assumption may be reasonable
for very simple IoT processor designs, but we will discuss this aspect further at the end of this section.) 
In contrast, for this particular \gemm shape,
the distribution of costs and the global execution time is highly dependent on the
algorithmic variant and micro-kernel dimensions. Thus, for this particular layer of MobileNetV1, both 
B3A2C0 and B3C2A0 tend to favor ``low-and-fat'' micro-kernels, such as $4 \times 24$, while
C3B2A0 yields better performance for ``squarish'' ones: 
$8 \times 12$ and
$12 \times 8$.

\begin{figure}[tbh!]
\centering % \vspace{0.5cm}
      \includegraphics[angle=-90,width=\textwidth]{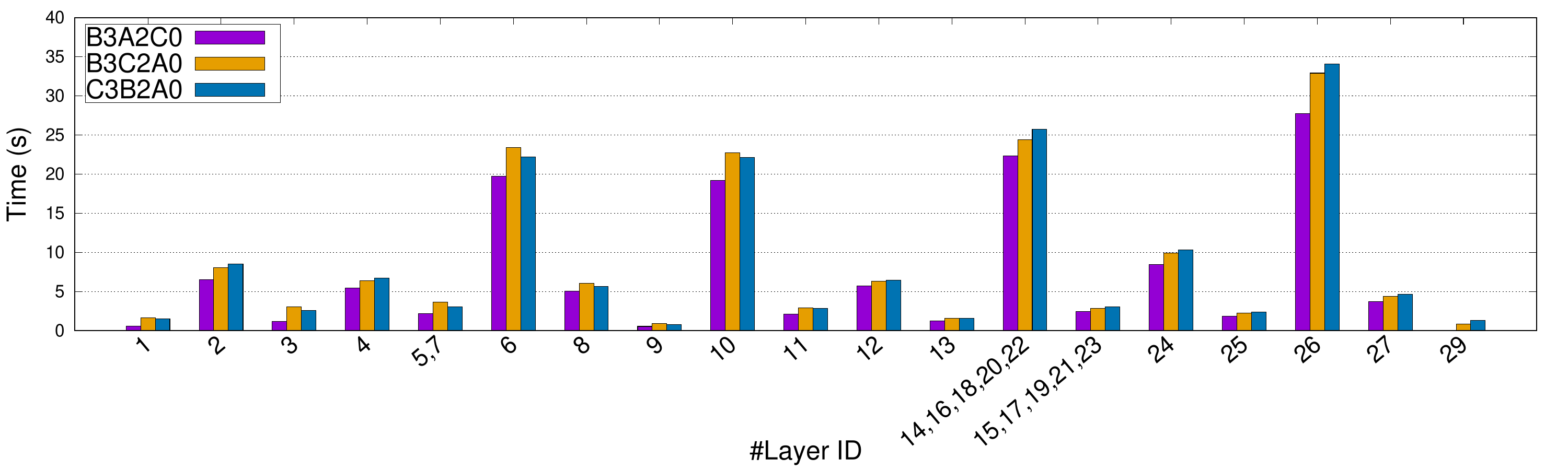}
\caption{Execution time of the three algorithms for the \gemm in MobileNetV1 
         estimated using the performance simulator calibrated for the GAP8.}
  \label{fig:mobilenet}
\end{figure}

Finally, Figure~\ref{fig:mobilenet} compares the estimated execution time
for the \gemm resulting from the application of lowering to all the convolution layers
of MobileNetV1. The particular dimensions of these layers are specified
in Table~\ref{tab:mobilenet}, together with the optimal micro-kernel dimension for each algorithmic variant
and layer dimensions.
(Layer \#28 is skipped because it does not correspond to a convolution operator.)

The results in this final experiment show that a high variability of the execution time, in accordance with the
heterogeneity of the \gemm shapes for the distinct layers, but also a general advantage of the 
B3A2C0 variant.
This was not totally unexpected as B3A2C0 mimics the baseline algorithm in BLAS instances such as those in GotoBLAS2, 
OpenBLAS and BLIS, and presents the advantage of reducing the number of stores in memory during the update of the
result $C$. However, we note that this variant depends on the underlying architecture offering an efficient
SIMD support for the outer product, which may not be the case for all Iot processors. For example, the GAP8
architecture is especially designed to deliver high performance for the scalar (or dot) product, which
favors the \gemm variants with $A$ resident in the processor registers
(C3B2A0 and B3C2A0). This would be reflected in a different (INT8) GOPS rates in our simulator,
depending on the type of micro-kernel and architecture design.
This architecture-specific adaptation of the simulator to the arithmetic units in the target processor is left as part of future work.

\begin{table}[htb!]
\begin{center}
\setlength{\tabcolsep}{6pt}
\begin{tabular}{|c||r|r|r|r|r|r|}
\hline
\#Layer ID & $m$ & $n$ & $k$    & B3A2C0      & C3B2A0      & B3C2A0
\\ \hline \hline
1         & 32  & 12544 & 27   & 4$\times$24 & 24$\times$4 & 8$\times$12 \\
2         & 32  & 12544 & 288  & 4$\times$24 & 8$\times$12 & 4$\times$24 \\
3         & 64  & 12544 & 32   & 4$\times$24 & 24$\times$4 & 12$\times$8 \\
4         & 64  & 3136  & 576  & 4$\times$24 & 12$\times$8 & 4$\times$24 \\
5,7       & 128 & 3136  & 128  & 4$\times$24 & 24$\times$4 & 4$\times$24 \\
6         & 128 & 3136  & 1152 & 4$\times$24 & 12$\times$8 & 4$\times$24 \\
%7         & 128 & 3136  & 128  & 4$\times$24 & 24$\times$4 & 4$\times$24 \\
8         & 128 & 784   & 1152 & 4$\times$24 & 12$\times$8 & 4$\times$24 \\
9         & 256 & 784   & 128  & 4$\times$24 & 24$\times$4 & 8$\times$12 \\
10        & 256 & 784   & 2304 & 4$\times$24 & 12$\times$8 & 4$\times$24 \\
11        & 256 & 784   & 256  & 4$\times$24 & 12$\times$8 & 4$\times$20 \\
12        & 256 & 196   & 2304 & 4$\times$24 & 12$\times$8 & 4$\times$24 \\
13        & 512 & 196   & 256  & 4$\times$24 & 24$\times$4 & 4$\times$24 \\
14,16,18,20,22 & 512 & 196   & 4608 & 4$\times$24 & 12$\times$8 & 4$\times$24 \\
15,17,19,21,23 & 512 & 196   & 512  & 4$\times$24 & 12$\times$8 & 4$\times$24 \\
%16        & 512 & 196   & 4608 & 4$\times$24 & 12$\times$8 & 4$\times$24 \\
%17        & 512 & 196   & 512  & 4$\times$24 & 12$\times$8 & 4$\times$24 \\
%18        & 512 & 196   & 4608 & 4$\times$24 & 12$\times$8 & 4$\times$24 \\
%19        & 512 & 196   & 512  & 4$\times$24 & 12$\times$8 & 4$\times$24 \\
%20        & 512 & 196   & 4608 & 4$\times$24 & 12$\times$8 & 4$\times$24 \\
%21        & 512 & 196   & 512  & 4$\times$24 & 12$\times$8 & 4$\times$24 \\
%22        & 512 & 196   & 4608 & 4$\times$24 & 12$\times$8 & 4$\times$24 \\
%23        & 512 & 196   & 512  & 4$\times$24 & 12$\times$8 & 4$\times$24 \\
24        & 512 & 49    & 4608 & 8$\times$12 & 12$\times$8 & 4$\times$24 \\
25        & 1024 & 49   & 512  & 8$\times$12 & 12$\times$8 & 4$\times$24 \\
26        & 1024 & 49   & 9216 & 8$\times$12 & 12$\times$8 & 4$\times$24 \\
27        & 1024 & 49   & 1024 & 8$\times$12 & 12$\times$8 & 4$\times$24 \\
29        & 1024 & 1000 & 1    & 4$\times$24 & 24$\times$4 & 24$\times$4 \\ 
\hline
\end{tabular}
\end{center}
\vspace*{-2ex}
\caption{\gemm operations in the convolution layers arising in MobileNetV1 transformed via lowering, and
         dimension of the optimal micro-kernel.}
\label{tab:mobilenet}         
\end{table}

\section{Discussion and Future Work}
\label{sec:remarks}

In order to address the heterogeneous zoo of IoT processor designs for edge computing, 
we have leveraged a performance simulator for estimating the execution costs of \gemm 
that offers very useful information about which algorithmic variant can better fit a particular architecture.

At the same time, we recognize this work needs to be
extended and improved along several paths. As part of future
work, we plan to explore several avenues:
\begin{itemize}
\item Micro-kernels with $A/B$ or $C$ resident in registers are usually
      cast in terms of distinct assembly SIMD (single instruction, multiple
      data) instructions. This needs to be taken into account in the calibration experiments.
\item Also, most current processors architectures are equipped with       
      DMA controllers. This complicates programming in order to orchestrate asynchronous transfers 
      with computation, and requires double buffering thus
      reducing the amount of memory for the buffers in the intermediate memory levels.
\item Finally, we plan to modify the memory model to take into account actual cache
      memories instead of scratchpads. This introduces challenges associated with
      modeling the effects of cache associativity, cache eviction, and 
      replacement policies.
\end{itemize}

%--------Acknowledments------------------------------------
\section*{Acknowledgments}
This work was supported by the research project
PID2020-113656RB-C22
of
MCIN/AEI/10.13039/501100011033.
%y por FEDER \textit{Una manera de hacer Europa}.
C. Ram\'irez is a ``Santiago Grisol\'ia'' fellow supported
by \textit{Generalitat Valenciana}.
Adri\'an Castell\'o is a FJC2019-039222-I fellow
supported by MCIN/AEI/10.13039/501100011033.
H. Mart\'inez is a ``Ayuda Postdoctoral'' fellow supported
by \textit{Consejería de Transformación Económica, Industria, Conocimiento y Universidades de la Junta de Andaluc\'ia}.

This project has received funding from the European High-Performance Computing Joint Undertaking (JU) under grant agreement No
955558. The JU receives support from the European Union’s Horizon 2020 research and innovation programme, 
and Spain, Germany, France, Italy, Poland, Switzerland, Norway.

\bibliographystyle{splncs04}
\bibliography{deep,hpc,biblio}
%\end{thebibliography}

\end{document}